# Two-step conversion of metal and metal oxide precursor films to 2D transition metal dichalcogenides and heterostructures

*Michael Altvater, Christopher Muratore, Michael Snure,\* Nicholas Glavin\**

M. Altvater, N. Glavin
Air Force Research Laboratory, Materials and Manufacturing Directorate, WPAFB, OH 45433 USA
Email Address: nicholas.glavin.1@us.af.mil

M. Altvater
UES Inc., Dayton, OH 45432 USA

C. Muratore
Department of Chemical and Materials Engineering, University of Dayton, OH 45409 USA

M. Snure
Air Force Research Laboratory, Sensors Directorate, WPAFB, OH 45433 USA
Email Address: michael.snure.1@us.af.mil

From the laboratory to real-world applications, synthesis of two dimensional (2D) materials requires modular techniques to control morphology, structure, chemistry, and the plethora of exciting properties arising from these nanoscale materials. In this review, we explore one of the many available synthesis techniques; the extremely versatile two-step conversion (2SC) method. The 2SC technique relies on deposition of a metal or metal oxide film, followed by reaction with a chalcogen vapor at an elevated temperature, converting the precursor film to a crystalline transition metal dichalcogenide (TMD). Herein, we consider the variables at each step of the 2SC process including the impact of the precursor film material and deposition technique, the influence of gas composition and temperature during conversion, as well as other factors controlling high quality 2D TMD synthesis. We feature the specific advantages to the 2SC approach including deposition on diverse substrates, low temperature processing, orientation control, and heterostructure synthesis, among others. Finally, emergent opportunities that take advantage of the 2SC approach are discussed to include next generation electronics, sensing, and optoelectronic devices as well as catalysis for energy-related applications; spotlighting the great potential of the 2SC technique.

**1. Introduction**

Two-dimensional transition metal dichalcogenides (TMDs) have attracted strong interest within the materials science and engineering communities due to promising electronic, optical, and catalytic properties exhibited by the layered structures.[1] The TMD atomic structure is composed of atomic layers wherein each layer comprises of a sheet of transition metals (M=Mo, W, Cr, Ti, Zr, Hf, V, Nb, Ta, Tc, Re, Pd, Pt, etc.) sandwiched by layers of chalcogen atoms (X=S, Se, Te), resulting in the chemical formulation $MX_2$ (e.g. $MoS_2$, $NbSe_2$, $WTe_2$). Within each layer, the transition metals form a triangular lattice covalently bonded to the chalcogen atoms typically either with a trigonal prismatic or octahedral coordination with select TMDs forming distorted versions of these lattices. While the TMDs are constructed by strong covalent intralayer bonds, subsequent layers are bound by relatively weak van der Waals interactions,[1,2] which results in unique properties including superlubricity,[3] high efficiency catalysis,[4,5] and advanced electronic/optical devices.[6]

A great reemergence of research interest in the van der Waals-stacked TMDs arose following the isolation and electronic characterization of graphene from graphite.[7] Shortly afterwards, it was realized that other layered materials, TMDs included, can be isolated in monolayer form[8] enabling the demonstration of high performing field-effect transistors and enhanced photoluminescence in $MoS_2$ monolayers.[6,9–11] Since these initial demonstrations, the growing family of 2D TMDs has shown to exhibit not just semiconducting properties, but a wide range of electronic states and behaviors. Early interest focused on the semiconducting TMDs such as $MoS_2$, $MoSe_2$, $WS_2$, and $WSe_2$, as they have a wide range of band gaps in the visible energy range. When thinned toward the monolayer limit, quantum confinement in TMDs can cause dramatic changes in band gap value and transitions between direct and indirect gaps. For example, the bandgap for MoS$_2$ expands from a 1.2 eV indirect gap in the bulk to a 1.8 eV direct gap in monolayer form.[12] In addition to layer number, band gaps can be finely tuned with strain, alloying, and doping[13–16] and exotic phenomena such as valley Hall effect and optical valley Zeeman effect have been observed in these monolayer materials.[17] The large family of 2D TMDs also include metals and semimetals which demonstrate exciting emergent electronic phases such as charge density wave, superconductivity, and Mott insulating phases.[18–23] Additional efforts to explore substitutional doping of magnetic atoms for transition metals in TMDs have even revealed ferromagnetic phases in doped TMDs.[24] Material properties can be tuned through post-synthesis doping, intercalation, and polymorph control. Further, their high surface-to-volume ratio and high compatibility with functionalization makes 2D TMDs exceptionally good platforms for sensing applications.

Despite numerous research-scale demonstrations showing promising device performance, TMDs have yet to find their place in industrial applications including as supplements to silicon-based electronics.[25,26] The same sensitivity that permits TMDs to such a wide range of applications also makes them inherently difficult to control their properties at wafer scale. The small thickness of 2D materials means that small variations in crystal structure, film thickness, interface quality, and electronic environment can lead to dramatic changes in electronic and optical performance. The superlative metrics of TMD performance are typically achieved through isolating single crystal thin TMD flakes from bulk crystals by cleaving them between van der Waals planes using adhesive tape. Tape-exfoliated flakes can range in thickness from hundreds of nanometers to single-layer thicknesses with lateral sizes of tens to hundreds of micrometers. The high degree of crystal quality and low defect concentrations achieved through flux zone, chemical vapor transport (CVT), and natural geologically grown bulk crystals has resulted in extremely high performing device demonstrations. Micromechanical exfoliation and transfer can produce the highest quality TMD devices, however, this technique is laborious and is impractical to scale to meet industrial needs. Other top-down approaches such as ball milling and liquid phase exfoliation are, in principle, more scalable, however, these methods only produce films with sub-micron grain sizes, limiting electronic performance. Similarly, more standard thin film deposition techniques such as sputtering, molecular beam epitaxy (MBE), and solution-based synthesis typically produce polycrystalline TMDs with grain sizes only reaching hundreds of nanometers. Limited grain sizes restrict the applicability of these films as grain boundaries and random crystallite orientations can dramatically degrade electronic and optical performance.[27] Chemical vapor deposition (CVD) techniques involving reaction of molecular precursors, including atomic layer deposition (ALD) and metal organic CVD (MOCVD), can achieve highly uniform, atomically thin, wafer-scale films; however, grain sizes are still small, production costs are high, and the chemical precursors are often toxic. Powder based CVD techniques can reliably achieve large grain sizes (some over hundreds of microns[28]) and relevant properties for many applications but wafer scale growth takes exceptionally long time and is still not commercially viable. In addition to TMDs finding an application space in modern electronics, TMDs have yet to find a physical location to be implemented alongside silicon-based electronics; potentially in the back end of the line (BEOL).[26]

Another synthesis approach, sometimes referred to as the two-step conversion (referred to as 2SC in this review) method, has emerged as a viable technique of producing wafer-scale TMD films. In this technique, a pure metal or metal oxide precursor film is first deposited using thin film deposition techniques (including MBE, sputtering, evaporation, pulsed laser deposition (PLD), or ALD, to name a few) followed by a second step where the deposited film is converted to a TMD by reacting with a vapor phase chalcogen source at elevated temperature (See Figure 1). The 2SC technique was first used to produce atomically thin TMDs in 2012 with the work of Zhan et al.[29] wherein the conversion of a ~1-5 nm e-beam evaporated Mo film was successfully converted to 2H-$MoS_2$ using elemental sulfur vapor. A similar work was reported that year creating $MoS_2$ via the 2SC method starting with a metal oxide precursor ($MoO_3$) with thicknesses between 0.8 and 3.6nm.[30] In both cases, HRTEM and Raman spectroscopy reveal the conversion results in one to few layer polycrystalline $MoS_2$ aligned parallel to the $SiO_2$ substrate. As the lateral dimensions of the TMD film are only limited by the size of the substrate and the thickness of the film is determined by the thickness of the precursor layer, this method was demonstrated to be a facile and scalable method of wafer-scale, atomically thin TMD production. Following the seminal work of Zhan et al., the 2SC method has been used to produce a wide range of ultrathin TMDs including $MoS_2$, $MoSe_2$, $MoTe_2$, $WS_2$, $WSe_2$, $WTe_2$, $PtS_2$, $PtSe_2$, $PtTe_2$, $TiSe_2$, $NbSe_2$, and heterostructures of the same.

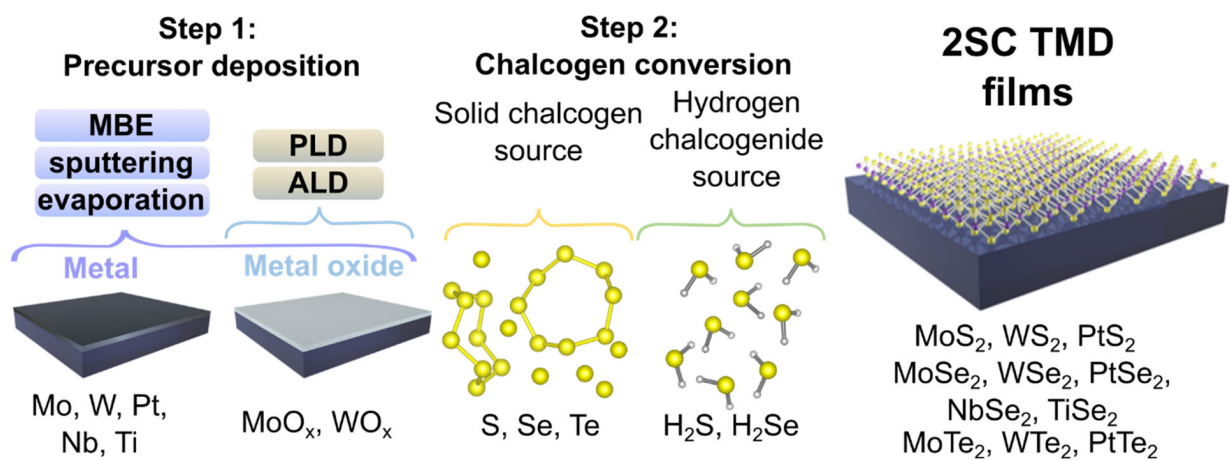

**Figure 1.** Schematic diagram of the 2SC process starting with precursor deposition (left) followed by conversion in chalcogen atmosphere (center) and a schematic of converted TMD films. Precursor sources, chalcogen sources, and resulting TMDs listed reflect those reported for 2SC growth to date.

In this review, we highlight the 2SC process and relative benefits as compare with other vapor phase deposition methods and provide key insights into the growth considerations and potential applications of this versatile synthesis technique. It is important to note that this technique has been referred to by several different names, which has caused some confusion in the community. The conversion of metal and metal oxide films to TMDs using a vapor-phase conversion process has many times been referred to as "CVD", "two-step CVD", and "thermally assisted conversion," to name a few.[31–35] In this review, we make the distinction between 2SC growth and other growth techniques, namely that the 2SC refers to a two-step growth process where a solid thin film (typically metal or metal oxide) is deposited on a substrate and then converted to a TMD through reaction with chalcogen vapors (Figure 2a).

## 2. 2SC within the 2D synthesis family

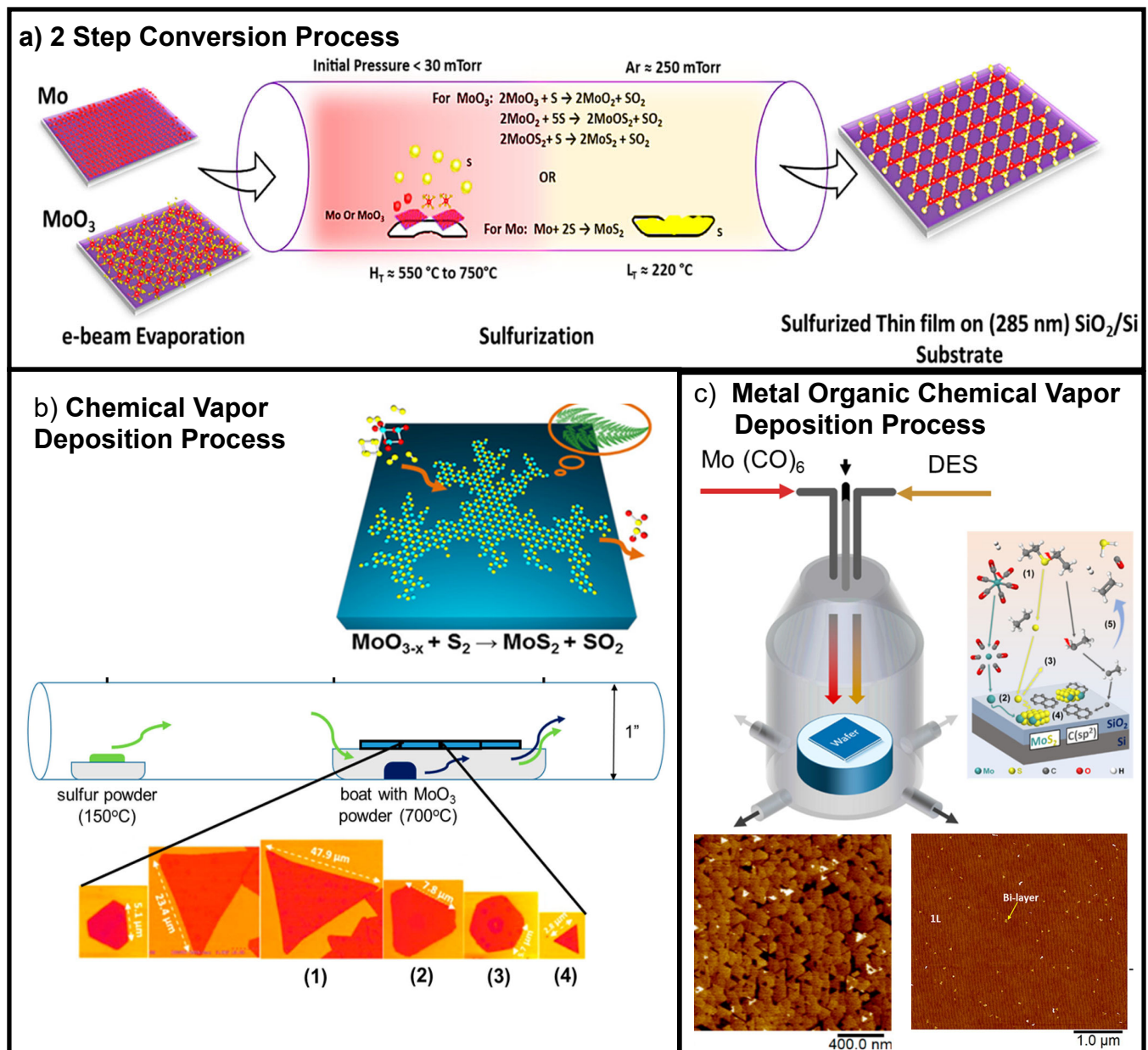

**Figure 2.** a) Schematic of the two-step conversion process; b) Schematic of the CVD growth process and position based variation in MoS2 morphology; c) Schematic of TMD MOCVD growth process and representative AFM images of WS$_2$ films during different stages of growth. a) Reprinted with permission.[36] Copyright 2023 American Chemical Society. b) Adapted with permission.[37,38] Copyright 2016 American Chemical Society. Copyright 2014 American Chemical Society. c) Adapted with permission.[39–41] Copyright 2017 American Chemical Society. Copyright 2021 American Chemical Society. Copyright 2021 American Chemical Society.

Chemical vapor deposition generally refers to a broad category of growth process covering a range of growth chamber configurations, chemistries, temperatures, and pressures. For growth of TMDs, CVD can be divided into a few common variations: 1) growth by sublimation of solid source precursors (Figure 2b) like M, W, $MoO_x$ or $WO_x$ and S, Se or Te,[42,43] or 2) MOCVD, which utilities metal organic (MO) sources for the TM and gas or organic sources for the chalcogen (Figure 2c)[44]. Even within these basic categories of CVD there is a great variation depending on the growth chamber, precursors, and process conditions. Generally, CVD variation 1), which will now be referred to as CVD for consistency with a broad adoption in the literature, is carried out in a hot wall horizontal tube furnace, with separate heating zones for the solid TM and chalcogen sources to control the vapor pressure of each. Source vapor is usually transported downstream using an inert gas, like Ar[45] or inert gas with a small amount of $H_2$ [46], to the heated growth substrate. On the other hand, MOCVD reactors and chemistries are more complicated where the partial pressure of MO TM source is regulated using independent control of the source temperature, pressure, and transport gas through a bubbler, and the chalcogen source can be either a gas phase precursor like $H_2S$ or an organic source like diethyl disulfide ($Et_2S_2$). Due to the wide range of process conditions and complexity of the TMD CVD process there is still a great deal which is unknown; however, both CVD and MOCVD basically follow a conventional nucleation and lateral growth process. The 2SC process is fundamentally different because the TM source (metal or oxide) is already on the substrate and not part of the vapor transport process, as will be discussed in the sections below. For CVD and MOCVD techniques, the nucleation step is critical to crystal quality, orientation, and grain size, which is controlled using process controls like- temperature, precursor partial pressure as well as substrate surface defects and step edges. Typically, the nucleation step occurs at lower temperature and high partial pressure to promote formation of stable islands. Followed by lateral growth to coalesce islands into a continuous film at higher temperature and lower partial pressure[34]. In these steps, the difference between the ability to control the partial pressure of precursors by CVD and MOCVD makes a significant difference. To control precursor vapor pressure of CVD solid sources temperature is the primary control knob, while in MOCVD precursor vapor pressures can be well controlled allowing for wider range of control over precursor partial pressure, metal to chalcogen ratios and temperatures. Figure 2b shows how the precursor partial pressure in the

CVD process can change dramatically as a function of distance from the source, resulting in dramatically different nuclei shape and size[37]. As a comparison, Figure 2c shows images of high density, well oriented, and uniform nuclei, which are coalesced into continuous monolayer films achieved by MOCVD. Generally, MOCVD nuclei tend to be smaller than CVD, which may result in high grain boundary and defect densities. Park et al. showed high quality $MoS_2$ with domain size up to 120 μm could be grown at lower temperature by MOCVD. Resulting in monolayer films with lower doping, lower strain, higher mobility, and narrower PL than achieved by CVD.[44] For electronic and optical applications, MOCVD grown TMDs tend to have electron mobilities of less than 20 $cm^2V^{-1}s^{-1}$ for $MoS_2$ and 10 $cm^2/Vs$ for $WSe_2$.[47–49] For more detailed reviews on CVD and MOCVD process we point the readers to references.[34,50–52]

Compared to the CVD and MOCVD processes, the 2SC process has numerous advantages including simplicity, reduced processing temperatures, material versatility, high throughput, ability to produce heterostructures, and complex patterned structures. Further, the 2SC method can produce TMD films of comparable quality and performance with CVD and MOCVD; as will be shown in later sections. However, the process does have limitations, which present challenges to controllably achieving device quality mono- to many-layer films and heterostructures. In the sections below, we review selection of the TM source, chalcogen source, and effect of process conditions on TMD film quality.

## 3. The 2SC Process
### 3.1. Transition Metal Precursor Deposition

Unlike conventional vapor deposition of TMDs, the 2SC method requires initial deposition of a solid-state transition metal thin film precursor. As typical transition metal diffusion is slow at the conversion temperatures used, the deposition of precursor material film defines the location and ultimate thickness of the converted TMD film using the 2SC method. This provides opportunities to use the precursor deposition step to pattern device structures, precisely control film thickness, employ flexible and unusual substrates, and conformably coat high-aspect ratio surfaces. Further, using a pre-deposited reactant makes the conversion process self-limiting while also providing a significant reduction in growth time compared to typical CVD by circumventing the known-issue of slow nucleation and growth rates of TMDs.

The low mobility of transition metals and slow crystallization rate of TMDs means that the physical and chemical properties of the precursor film can have a large impact on the properties of the converted TMD film. In this section, we give an overview of mature technologies used to deposit ultra-thin metal and metal oxide films including atomic layer deposition (ALD), pulsed laser deposition (PLD), molecular beam epitaxy (MBE), thermal and electron beam evaporation, and magnetron sputtering while highlighting the key strengths of each technique and typical properties of TMD films converted using them.

2SC synthesis of 2D TMD films has been demonstrated using a variety of precursor phases include hydrothermal metal oxides,[53] sublimation of metal foil in air,[54] growth on metal foil,[55] and printing solutions of sodium molybdate,[56] ammonium thiomolybdate,[57] ammonium metatungstate,[56,58,59] or ammonium heptamolybdate tetrahydrate.[60] These methods, being relatively new, lack technological maturity for repeatable application of 2D materials over large areas with the exquisite thickness control necessary to ensure uniform properties.[61–63]

### 3.1.1. Comparison of metal precursor application methods

ALD of metals and metal oxide materials remains one of the industry standard fabrication techniques for exquisite thickness control, conformal deposition, and low temperature synthesis. 2-step TMD synthesis utilizes ALD's strengths of atomic layer control, conformal deposition on high aspect ratio surfaces, and high purity,[64] thereby eliminating barriers from limited options for metal-chalcogen precursors. Materials from ALD metal oxide precursors with high charge mobility and on/off ratios are found in the literature, but typically require high chalcogenization transformation temperatures (500-1000 °C).[65–68] Reports on materials synthesized <500 °C are possible with converted ALD films; however, high temperature conversion appears to be a requirement for useful device properties.[69–71] Plasma-enhanced variations of ALD where precursor gases are excited and/or dissociated prior to interactions with the substrate, reduce metal oxide film formation temperature but have no impact on precursor transformation temperature from oxide to chalcogenide.[72] One attractive feature of ALD is facile selective area deposition[73] for patterning of materials with different characteristics (e.g., p-type/n-type), which can potentially result in simplified device fabrication and integration. While a few precursor films prepared by ALD have been demonstrated (MoO$_3$,[69,72,74–80] WO$_3$,[65,72,75,81,82] Pt[83]), 2SC precursor materials are limited to the available ALD chemical precursor options which might hinder the use of ALD for the production of certain TMDs.[66]

Pulsed laser deposition (PLD) relies on laser energy (typically UV at 248 nm) for irradiation of electrically insulating solid target materials to induce vaporization, thereby enabling direct deposition of thin films. For PLD (and each of the other PVD processes listed below), high purity solid source materials as high as 99.95%, and gas phase components with purity up to 99.9999 percent are used. The process is better suited for metal oxides than metals as the higher surface reflectivity and thermal conductivity of metals reduces energy absorption and rapid diffusion of heat from the laser pulse into the target material. In general, scalability and repeatability of PLD processes can be challenging as film properties depend upon the oblique incidence of vaporized material.[84] Additionally, as the target erodes away and the interaction of the incident light with the surface is altered due to chemical, morphological, and dimensional changes with each incident laser pulse.[84] Despite scalability challenges, PLD is an appealing option for 'material prototyping' for combinations of mixed oxide phases and materials compositions.[85] For the production of 2SC TMDs, PLD has been employed to produce metastable and sub-stoichiometric oxide precursors, such as MoO$_2$[86] and WO$_{3-x}$,[87] which tend to promote conversion. The high kinetic energy of incident film constituents can facilitate low-temperature direct crystalline metal oxide deposition.[88]

Evaporation of refractory metals such as Mo and W with high sublimation temperatures requires specialty evaporation systems with high power delivery to source material and management of residual thermal energy. Sublimation of transition metal or metal oxide powder, however, occurs <750 ºC which is easily attained in conventional evaporation systems. Evaporation of metal oxides MoO$_3$ is therefore a common practice for 2D synthesis with ease of control over the deposition parameters by regulating the temperature and flow of carrier gas such as argon and/or oxygen. Substrate temperatures for growth via evaporation range from 400–600 °C, with reports of larger layered crystals around 550 °C.[89–92]

Molecular beam epitaxy (MBE) is an evaporation process modified via utilization of a Knudsen cell,[73] measuring the vapor pressure of the evaporating substance in real time to quantify evaporated material and meter out selected fluxes to condense on the growth substrate. MBE therefore allows controlled deposition rates of metals/metal oxides to easily achieve sub-monolayer films. Directly growing TMDs by MBE, however, requires large chalcogen pressures[93,94] introducing chalcogen contamination in the UHV chamber, requiring a system dedicated for growth of one type of chalcogen TMD (i.e., S or Se). Additionally, specialized vapor cracking Knudsen cells, such as hot-wall cells, are often required to decompose chalcogen clusters.[95] Sulfur-containing TMDs have successfully been used as sulfur sources in MBE as the vapor pressure of the chalcogen is much higher than the metal, thus a concentrated chalcogen flux evolves with little metal contamination.[96] A key differentiator of MBE over other processes considered here is the strict growth substrate selection and preparation requirements. Often materials with small (<3%) lattice mismatch are selected as substrates for traditional MBE processes, however van der Waals materials are more accommodating of interfacial strain due to lattice-mismatch. Treatment of the substrate surface to remove oxygen and expose necessary surface termination is typical for MBE. Common MBE substrates for 2SC growth include highly oriented pyrolytic graphite (HOPG),[97] sapphire,[97] noble metal single crystals (111 surface),[98] complex metal oxides,[99] and bulk oriented TMDs.[100] MBE processes are conducted under very clean UHV conditions, enabling deposition of high purity materials. Additionally, MBE systems are often equipped with surface analysis tools, such as reflection high energy electron diffraction (RHEED)[101] and Auger spectroscopy systems[102] for in situ analysis during growth or between growth steps, enabling studies of fundamental growth mechanisms.

As sputtering is one of the most industry-standard processes used to deposit metal and metal oxide films, it benefits from large area, highly customizable growth. Also, conformality of sputtering,[98] especially for processes engineered to have large fractions of ionized flux,[103] more closely matches that of ALD than other line-of-sight PLD and evaporation deposition approaches. The approach is well suited for effective large-area growth of pure metal or metal oxide 2D TMD precursors. Sputtering typically takes place in an evacuated chamber backfilled with ultra-high purity inert and/or reactive gas. A 'target' of pure metal or an oxide may be sputtered, or slowly eroded from impacts with gas ions with incident kinetic energies >100 eV.

The flux of material from a compound target is often approximately the same as the composition of the target itself due to balances in sputter yields and target surface composition, although gas-phase scattering processes and differences in angular and energy distributions between species ejected from the target may impact flux of sputtered material to the growing film.[104] Additionally, the film grown from this flux to the substrate may be sub-stoichiometric due to differences in sticking coefficient of incident components. Studies of sputtered metal and metal oxide films reveal lower TMD transformation temperatures than for ALD, and larger grain sizes and S:Mo ratios for transition metal oxide films compared to pure transition metals.[105,106] The technique is not just limited to synthesis of monolithic films, as there are reports of sputtered stacks of alternating metals to be later chalcogenized in one step.[61,62,73] One of the primary advantages of magnetron sputtering of TMDs and precursor films is the exquisite control of film growth kinetics by controlling the energy of the incident material flux using tools such as modulation of power to the sputter target and applied magnetic fields.[107–109] Additionally, similar to MBE, sputtering is often conducted in UHV chambers for high purity materials and clean interfaces.

A survey of available methods reveal the choice of precursor deposition method (ALD or PVD), precursor properties including composition, thickness, continuity,[97] roughness, extent of crystallinity, processing conditions, and even unexpected parameters such as water vapor exposure during transformation[110] have a strong impact on the characteristics of the transformed TMD material. These characteristics include continuity, morphology, and defect density of the chalcogenized materials. A few studies directly correlate precursor oxide and TMD properties,[53,111,112] providing enough data to show the composition and morphology of substrate is critical as propagation of features on the surface will be amplified via kinetic roughening, which is a key element of 2D materials continuity and ordering. Additional studies are needed to realize large area materials with atomic-scale ordering matching that of mechanically exfoliated materials. With growing numbers of published reports on structural-dependence of properties of 2D metal oxide films[98,113,114] such as $MoO_3$,[90,115,116] $WO_3$,[117–119] and other transition metal oxides, collaboration between communities with sharing of engineered oxides for subsequent conversion to TMDs[87,120] will expand the range of 2D TMD materials properties and applications.

## 3.2. Vapor-based conversion/ gas phase conversion

### 3.2.1. Effect of Metal vs Oxide on conversion

Many of the advantages of the two-step process result from the use of thin metal or metal oxide films on a substrate as the TM source. By using thin film precursors (Section 3.1), the challenges discussed above with forming continuous TMD films by CVD and MOCVD through nucleation and coalescence are avoided. Although ultra-thin (≤1 nm) continuous metal and metal oxide films can be deposited, it is important to understand the advantages of both sources. First off, for some TMDs, metal oxides are not a suitable TM source (Pt, Hf, Zr or Ti) due to oxide stability or lack of oxide, as such we will primarily focus on TM precursors for Mo and W based TMDs.

Metal Mo and W sources have higher melting temperatures and lower vapor pressure than metal oxide sources but have higher metal density and are less reactive. For the CVD process, a high vapor pressure TM source is desirable as it allows for effective mass transport, for the 2SC process it can lead to inhomogeneous or discontinuous TMD films due to precursor desorption. The low vapor pressure of Mo and W metal sources allows for the thickness of the TMD to be set by the well-controlled metal deposition process. The high melting temperatures of Mo (2623 ºC) and W (3422 ºC), linked to their low vapor pressures, limits grain growth due to the comparatively low TMD conversion temperature (< 1000 ºC). In work comparing, $WSe_2$ grain size formed by selenization of W and $WO_3$ using Se vapor, metal W films resulted in a grain size of ~0.5 um while $WO_3$ resulted in grain sizes as large as 1.5 um (Figure 3a). However, $WSe_2$ films from $WO_3$ had a larger variation in grain size due to higher volatility of $WO_3$.[121] Photodetectors showed better performance for the devices fabricated from $WSe_2$ films using $WO_3$ than the W source due to lower defect density. The TM atom density of the precursor films can determine the minimum TMD thickness achievable. Conversion of $MoO_3$ to $MoTe_2$ using Te vapor results in an 8x smaller expansion than Mo, due to the much lower density of $MoO_3$.[78] It was also found that both the semiconducting 2H and semimetallic 1T' phases of $MoTe_2$ could be formed from $MoO_3$, while only the 2H phase was formed using Mo.[122] In a report by Fatima et al.[36] they found sulfurization of 0.5 nm thick Mo films resulted in few layer (~3 nm) thick $MoS_2$, while monolayer $MoS_2$ was formed from 0.5nm thick $MoO_3$ (Figure 3b) with monolayer films showing strong PL emission.  Additionally, films formed from MoO3 had p-type doping, high hole mobility, and showed resistive switching due to sulfur vacancies.[36]

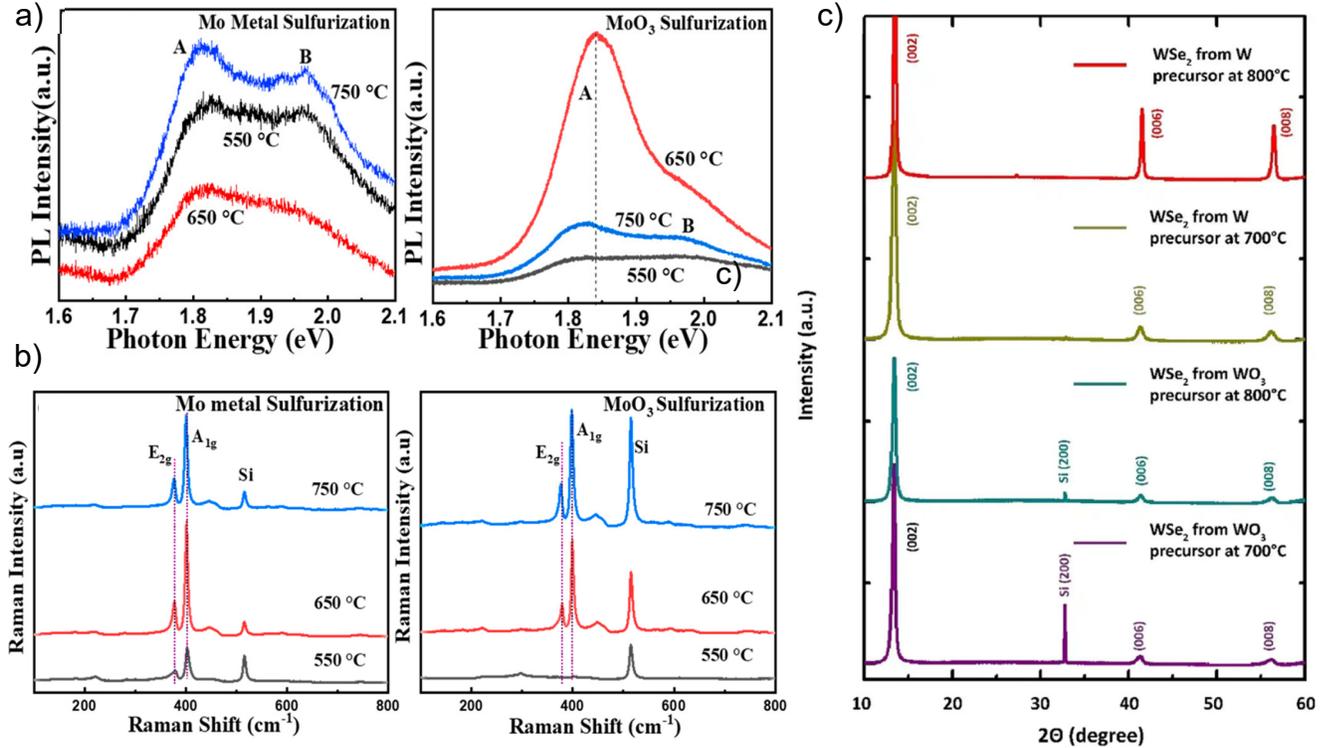

**Figure 3.** Comparison of a) PL and b) Raman spectra from MoS2 films formed by sulfurization of Mo and MoO3 films; c) XRD scans from WSe2 films formed by selenization of W and WO3. a,b) Reproduced with permission.[36] c) Reproduced with permission.[121] Copyright 2022 Elsevier.

The conversion to TMDs from metals and metal oxides is quite different and will also depend on the chalcogen source as discussed in the next section. Reaction of a metal sources with vapor from an elemental source, we can consider a simple two element reaction.  However, to form a TMD from a metal oxide is a more complex process involving three or more (TM, chalcogen, oxygen, hydrogen) components interacting with a required step to replace the oxygen and sweep reaction by-products away. Formation of volatile sulfur oxide species using an excess of S vapor can sufficiently replace and remove oxygen.  With insufficient sulfur, oxysulfides may form reducing materials quality.[45,123]

Control of the temperature ramp rate during exposure of metal and metal oxide film to chalcogen vapor plays a significant role in the crystallization of MoS$_2$.[69,74,76] For identical precursor films chalcogenized at high and low temperature ramp rates, there is competition of crystal growth from the amorphous MoS$_x$ matrix and nucleation of new crystal domains in that matrix. The slower heating rate may suppress the nucleation rate, allowing crystalline MoS$_2$ domains to grow larger before coalescing with other domains. This correlation may also be related to competition between sublimating precursor phases and sub-stoichiometric phases (i.e., MoO$_2$) which grow rather than sublime, suggesting the precursor oxidation state can have dramatic effect on the converted TMD material as well.[109]

### 3.2.2. Chalcogen Sources

A variety of chalcogen sources can be used for the 2SC synthesis process, which can impact the quality, morphology, and thickness of the resulting TMD films. From the early demonstrations of this process, solid chalcogen sources have been the most widely used due to simplicity, cost, availability, and safety. Solid sources require as little as an evaporation boat, carrier gas and a heated zone to create a suitable reactive environment for TMD conversion of metal or oxide precursors. However, the chalcogen vapor pressure from solid sources is dependent on temperature, pressure, carrier gas, and surface area making it difficult to control.[112,124,125] This commonly used powder vaporization route may not be suitable to supply a continuous and constant source of precursors necessary to achieve the scale of commercial semiconductor manufacturing processes. Comparatively, the partial pressures of gas sources for chalcogens are easy to control and are high purity, but are more toxic and highly reactive. Hydrogen chalcogenide (H$_2$S and H$_2$Se) sources have the added advantage of stability and simple pyrolysis products.[126] Hydrogen telluride is not available due to its poor stability. Evaporation of Se from solid source is known to produce vapor containing multiple Se clusters in addition to Se$_2$, which additionally complicates the reaction chemistry.[127]

Compared to S, Se and Te have a much weaker chemical activity and lower vapor pressure presenting obstacles to synthesis of selenide and telluride based TMDs from elemental precursors.[128] Sulfur has a vapor pressure reaching nearly 1 atm at 500 °C compared to 0.5 atm for Se and 0.001 atm for Te at 500 °C[32,129] Sulfur has been shown to efficiently react with metal or metal oxide sources as described above. Its reactions on oxidized surfaces readily forms volatile $SO_2$ to remove oxygen, while $SeO_2$ and $TeO_2$ are solids. As such the introduction of $H_2$ has been shown to promote the formation of selenides and tellurides. Lin *et al.*[130] showed selenization of $WO_x$ from Se vapor required the presence of $H_2$. In pure $N_2$, $WO_x$ films did not convert, while films selenized using a $H_2$:$N_2$ flow of 100 sccm/20 sccm concentration had the highest conversion ratio and narrowest Raman peaks. Both lower and higher $H_2$ flow rates resulted in similar conversion rates, as determined by XPS, but had broader Raman peaks due to oxygen impurities, selenium vacancies, and other defects. Growth of $MoTe_2$, by Hynke *et al.*[78], from $MoO_x$ and Te reported $H_2$ was required. Tellurization of Mo and W metal films was found to be strongly dependent on the presence of $H_2$.[131] Mo and W films tellurized in pure $N_2$ did not form any detectable $MoTe_2$ or $WTe_2$ at 650 °C, while the introduction of only 5% $H_2$ in $N_2$ did. By increasing the $H_2$ concentration to 100%, the best quality $MoTe_2$ and $WTe_2$ were realized. Due the high chemical activity of Pt it was show that $PtTe_2$ could be formed without $H_2$;[132] however, $H_2$ significantly improved crystallinity of $PtTe_2$ as determined by Raman and X-ray diffraction.[131] In reactions with metal oxides, the $H_2$ can accelerate the reaction through formation of $H_2O$. Selenide and telluride based TMDs have high Gibbs energy (compared to sulfides) creating a significant challenge to synthesis through direct reaction. By introducing $H_2$, it can form hydrogen chalcogenides, which are more reactive than the elemental vapor.[133]

Compared to elemental sources the chemistry from hydrogen chalcogenides is different due to availability of atomic hydrogen. If the growth temperature is high enough and a hot wall furnace is used, some amount of gas pyrolysis would be expected to form chalcogen vapor and $H_2$. However, when considering that only 50% of $H_2S$ will decompose to $H_2$ and $S_2$ at 1050 °C[134] and only 50% $H_2Se$ into $H_2$ and $Se_2$ at 500 °C,[135] surface reactions are likely to play a role during precursor conversion. In fact, high quality $MoSe_2$, $WSe_2$ and $MoSe_2$/$WSe_2$ superlattices were formed through selenization of Mo and W films at temperatures below 500 °C due to the reactivity of $H_2Se$.[61] This can also simplify the reaction chemistry with metal oxides to a two component $MoO_x + 2H_2S \rightarrow XH_2O + MoS_2$ vs. a three component $MoO_x + H_2 + S_2 \rightarrow XH_2O + MoS_2$ reaction. However, it has been shown that if the $H_2S$ concentration is too high the $MoS_2$ quality will degrade due to the presence of strongly reducing H.[136] Okada *et al.* reported that as the $H_2S$ flow was increased from 0.5 to 2 sccm the $WS_2$ PL was improved due to reduction of S vacancies. However, when the H₂S flow was increased to 4 sccm a significant decrease in PL was observed because of H induced etching.[137]

Alternative chalcogen sources like organic dimethyl selenide (DMSe) and plasma sources have also been demonstrated. Organo-chalcogen sources are widely used for MOCVD growth of TMDs offering precise vapor pressure control and are considerably safer than hydride sources. Two-step growth of large area WSe₂ growth has been demonstrated from reaction of a WO₃ film with (DMSe). The DMSe source decomposes into hydrocarbons and Se for Se-O exchange. Compared to films formed using elemental Se, no appreciable difference in material quality was observed.[138] For synthesis of crystalline TMDs directly on flexible polyimide substrates at 250 °C, selenization using a Se plasma source was demonstrated.[139]

## 4. Quality and Performance Characterization

In general, any technique used to synthesize 2D materials must focus on the resulting material performance and the subsequent scalability and reliability necessary for the particular application of interest. A well-controlled process should demonstrate wafer-scale uniformity of film properties while simultaneously maintaining microscopic-scale morphology control; including layer number control and film coalescence. Control of crystallinity and defect concentrations is required for optimizing films for different applications. For example, high crystallinity and low defect concentrations are desired for high electronic performance, while the opposite can be desired for catalytic or sensing performance (or, at least, some interplay between the two). The resulting 2SC TMD films should then also exhibit high compositional and phase purity. Below, standard film characterization techniques are shown to provide feedback on the 2SC process required for process optimization which can realize state-of-the-art 2SC TMD films with wafer-scale, highly crystalline, phase pure, low-defect, atomically thin TMD films.

One of the intrinsic benefits of the 2SC process is the modularity, reliability, and industry-relevant scalability of the process. As the thickness is governed by the precursor film, the 2SC method admits a wide range of thicknesses spanning from monolayer to 10's of nanometers shown in Figure 4a.[140] Additionally, the areal coverage is also dictated by the initial precursor film, which can be visibly observed in Figure 4b or even patterned to create more complicated film structures.[130] In the case of wafer-scale growth of WSe2, Raman mapping across multiple spots of the sample (shown in Figure 4b) clearly indicates structural uniformity across the wafer on several length scales with consistency of the Raman peak positions, intensity, and FWHM.[130] While the precursor deposition step dictates the thickness and uniformity of 2SC thin films, the temperature of the conversion step plays a key role in the final chemistry and structure. At elevated temperatures, the conversion process has shown to result in complete conversion to TMDs with little oxygen incorporation and optimal stoichiometry, as indicated by Figure 4d and 4e,[130] and improved crystal structure and grain size (Figure 4f).[141]

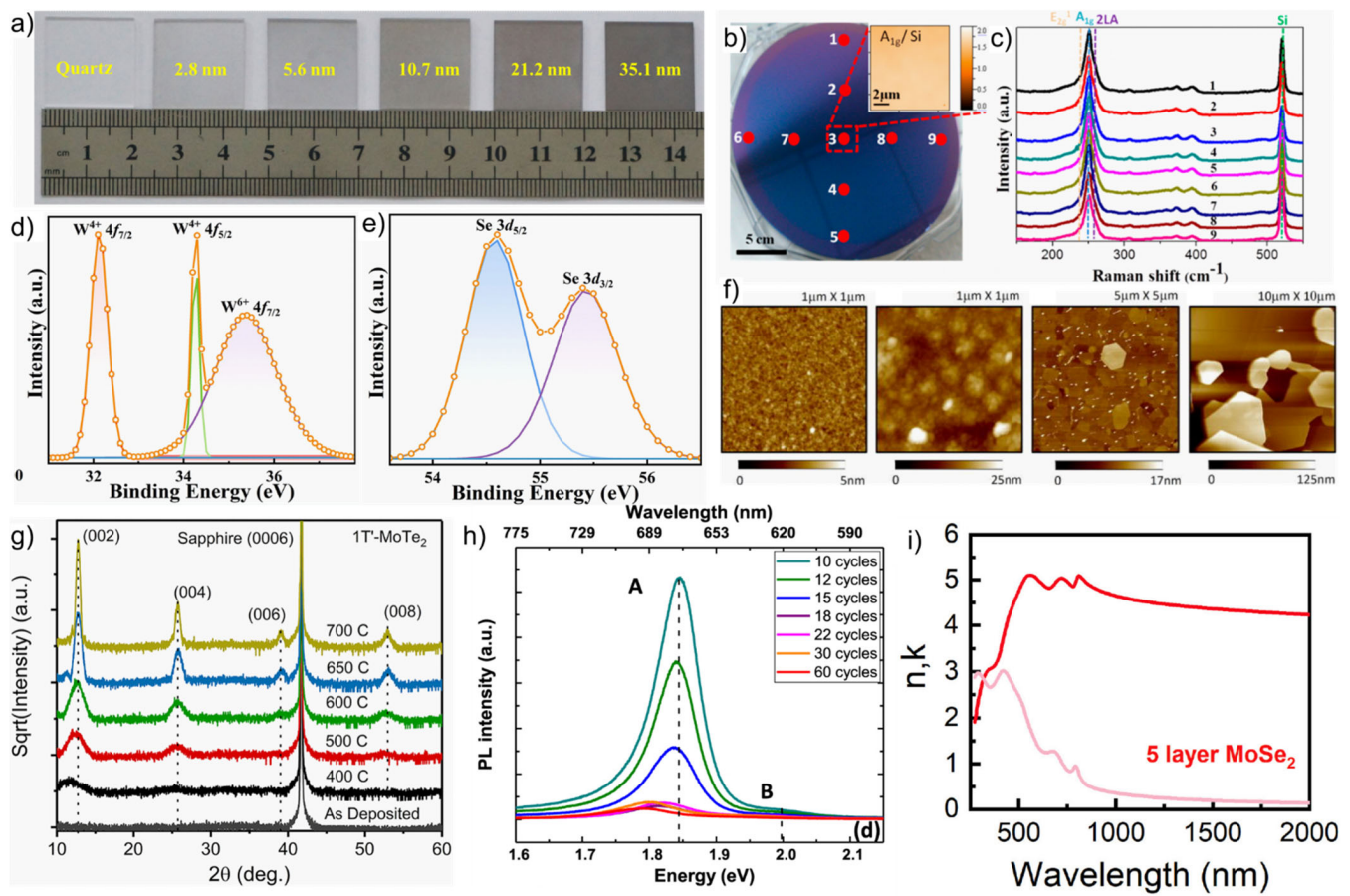

**Figure 4.** a) Optical image of 2SC WSe2 films of varying thicknesses on quartz with ruler for scale; b) Optical image of wafer-scale 2SC WSe2 on a 4 inch SiO2 wafer; c) Raman spectra obtained at several locations across the WSe2 film shown in (b); d) High resolution XPS spectra centered at the W 4f and e) Se 3d core transitions with no indication of oxide after conversion; f) AFM topography maps of MoS2 films converted at temperatures from 500 °C (left) to 1100 °C (right); g) XRD patterns obtained from MoTe2 films converted at different temperatures; h) PL

spectra of 2SC MoS$_2$ films converted from ALD MoO$_3$ films of varying thickness; i) Real and imaginary parts of the complex optical index of a 5-layer MoSe$_2$ film obtained through fitting VASE spectra. a) Reproduced with permission.[140] Copyright 2023 American Chemical Society. b,c,d,e) Reproduced with permission.[130] Copyright 2018 American Chemical Society. f) Reproduced with permission.[141] Copyright 2013 AIP Publishing g) Reproduced with permission.[131] Copyright 2022 AIP Publishing. h) Reproduced under terms of the CC-BY license.[72] Copyright 2020 IOP Publishing Ltd. i) Adapted with permission.[61] Copyright 2023 Elsevier.

With proper process control to obtain high crystal quality and purity, the optical performance of 2SC TMD films is competitive with other standard growth methods. In TMD thin films, photoluminescence is difficult to observe except in the case of semiconducting monolayers. It is often found that PL is quenched, not only due to changes in the material band structure but also due to defects, small grain sizes, and interactions with the substrate (including charge traps, chemical bonding, and unreacted metal at the substrate-TMD interface). Figure 4h gives an example of the PL spectra from MoS$_2$ converted from multiple cycles of ALD MoO$_3$. The observation of strong PL at and above one layer thickness is a testament to the quality of the films able to be produced by the 2SC method.[72] With variable angle spectroscopic ellipsometry (VASE), Motala et al.[61] demonstrate that 2SC films of MoSe$_2$ and WSe$_2$ are well suited as optical coatings exhibiting a refractive index greater than 4 across the optical spectrum making these films excellent optical trapping elements for optoelectronics (Figure 4d).

The ultimate structural characterization technique for TMD thin films, especially grown by the 2SC approach, is transmission electron microscopy (TEM). TEM is a powerful technique to investigate the crystal structure of 2SC TMD films with real-space atomic resolution. Figure 5a-d shows atomically resolved TEM images of a 2SC NbSe$_2$ film identifying grain sizes >20 nm across, high crystallinity, structural defect types, polymorph, and stacking control in one measurement.[63] Atomic layer control is demonstrated using cross-sectional TEM images of a 2SC MoS$_2$ film in Figure 5e and 5f.[142] Further, TEM can typically be coupled with energy dispersive spectroscopy (EDS), an X-ray spectroscopic technique, to distinguish elemental species within the TEM field of view. The combination provides atom-by-atom chemical mapping by using x-rays emitted by core electron state transitions as an element-specific marker. This is used to provide valuable feedback on the 2SC conversion process as demonstrated by Wu et al. in Figure 5g and 5h which identifies MoS$_2$ layers growing around molybdenum oxide clusters for films converted in very low chalcogen concentrations.[142]

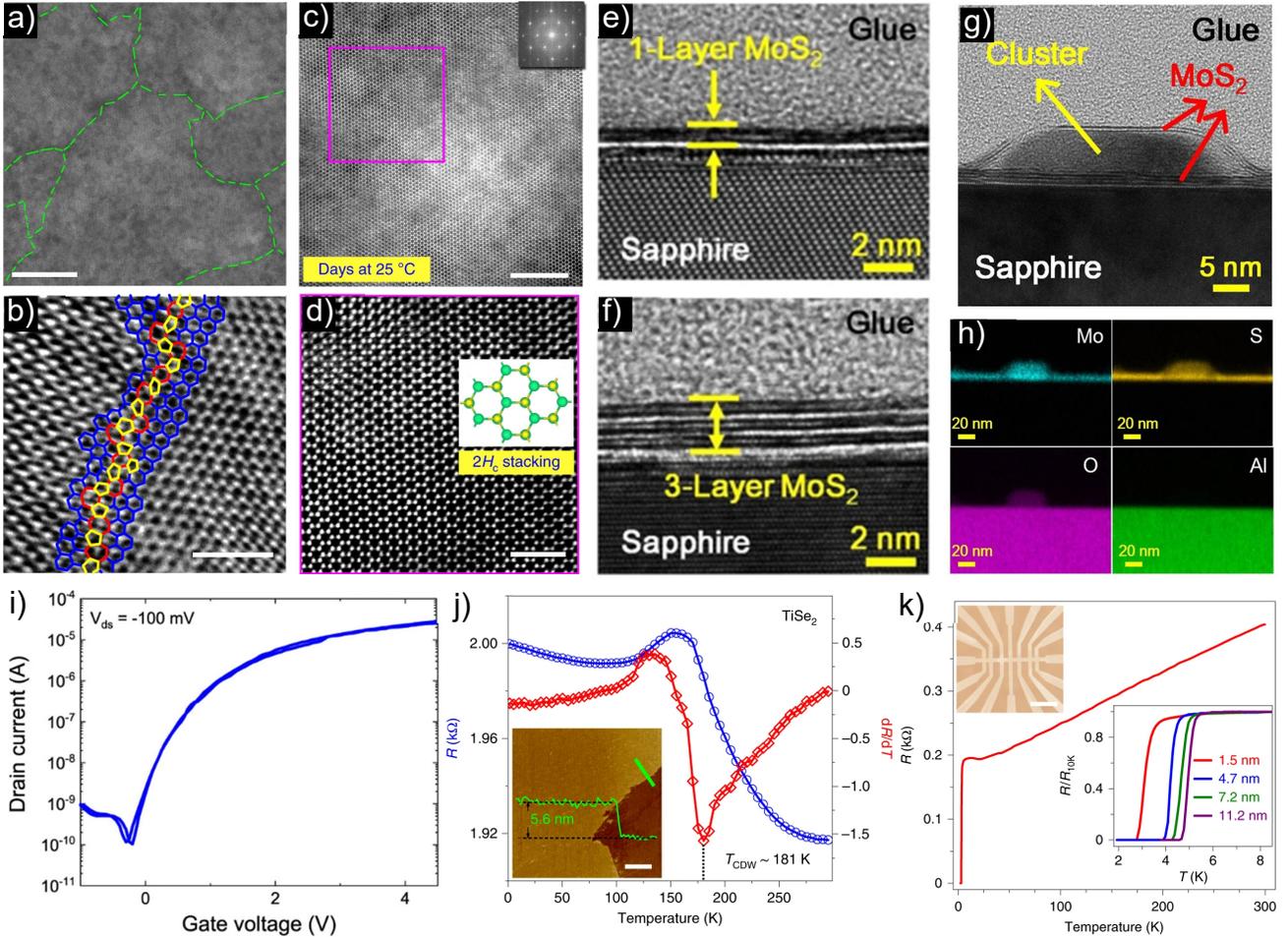

**Figure 5.** a) Large area and b) zoomed in TEM image at atomic resolution showing grain size and grain boundary structure in a 2SC NbSe$_2$ film c) Large area and d) zoomed in TEM image displaying lattice structure and stacking order in 2H-NbSe$_2$. Scale bars are 20 nm (a), 2 nm (b), 5 nm (c), 5 nm (d); e) Cross-sectional TEM of 1-layer MoS$_2$ and f) 3-layer MoS$_2$; g) A TEM image and h) EDS elemental maps of MoS$_2$ that has grown around a molybdenum oxide cluster; i) Electronic performance of a MoS$_2$ transistor; j) Temperature-dependent resistance of a 2SC TiSe$_2$ film and AFM topography (inset) showing a CDW transition at 181K; k) Temperature-dependent resistance measurement of a 2SC NbSe$_2$ film showing a superconducting transition between 3-5K depending on film thickness. a,b,c,d) Reproduced with permission.[63] Copyright

2019 Spring Nature. e,f,g,h) Reproduced under terms of the CC-BY license.[142] Copyright 2017 Springer Nature. i) Reproduced under terms of the CC-BY license.[80] Copyright 2023 Springer Nature. j,k) Reproduced with permission.[63] Copyright 2019 Springer Nature.

Most of the above characterization techniques primarily focus on directly obtaining microscopic structural and chemical properties in TMD films. Room temperature transport characterization of 2SC TMD films has primarily focused on the transistor performance metrics of the semiconducting TMDs. 2SC films can be either grown or transferred onto $SiO_2$ on Si wafers to produce field effect transistor (FET) devices. A simple measurement of current through the film versus gate voltage can provide measures of field-effect mobility, ON/OFF current ratio, subthreshold swing, and charge trap densities which determine the ultimate performance ability of a FET device. While the mobilities for the TMDs have been theoretically predicted[143] and experimentally observed in high quality, single crystal devices to be in the hundreds of $cm^2V^{-1}s^{-1}$,[9] the current state of the art 2SC films have demonstrated promising mobilities up to a few tens of $cm^2V^{-1}s^{-1}$ (Figure 5i),[80,86] leaving room for improvements to crystal quality and uniformity. Temperature dependent resistivity measurements of metallic (and semi-metallic) TMD films can reveal information about defect scattering in defective films[112,124] but in pristine films of certain TMDs, one can observe the electronic effect of structural phase transitions, such as the 1T' to 1Td phase transition in $MoTe_2$,[131] as well as correlated electronic phase transitions, such as the charge density wave (CDW) phase transition in 2SC $2H-TiSe_2$ (see Figure 5j) or the superconducting phase transition in 2SC $2H-NbSe_2$ (see Figure 5k).[63]

## 5. Orientation Control

In two-dimensional materials, the orientation of the crystallographic layers plays an important role in energy transport, intercalation, and stability, among other notable properties. In general, the 2SC process has been demonstrated as one of the more versatile synthesis techniques in controlling the layer orientation as it has two distinct growth modes under differing conversion conditions. The TMD layers can grow horizontally, with their basal planes parallel to the substrate, designated Type II or horizontal growth, or they can align vertically, perpendicular to the substrate, designated as Type I (Figure 6a).[144] It has been noted in several reports that one can transition between horizontal and vertical growth by controlling the precursor film thickness,[126,145] gas species used in the conversion step,[126] and annealing conditions.[146]

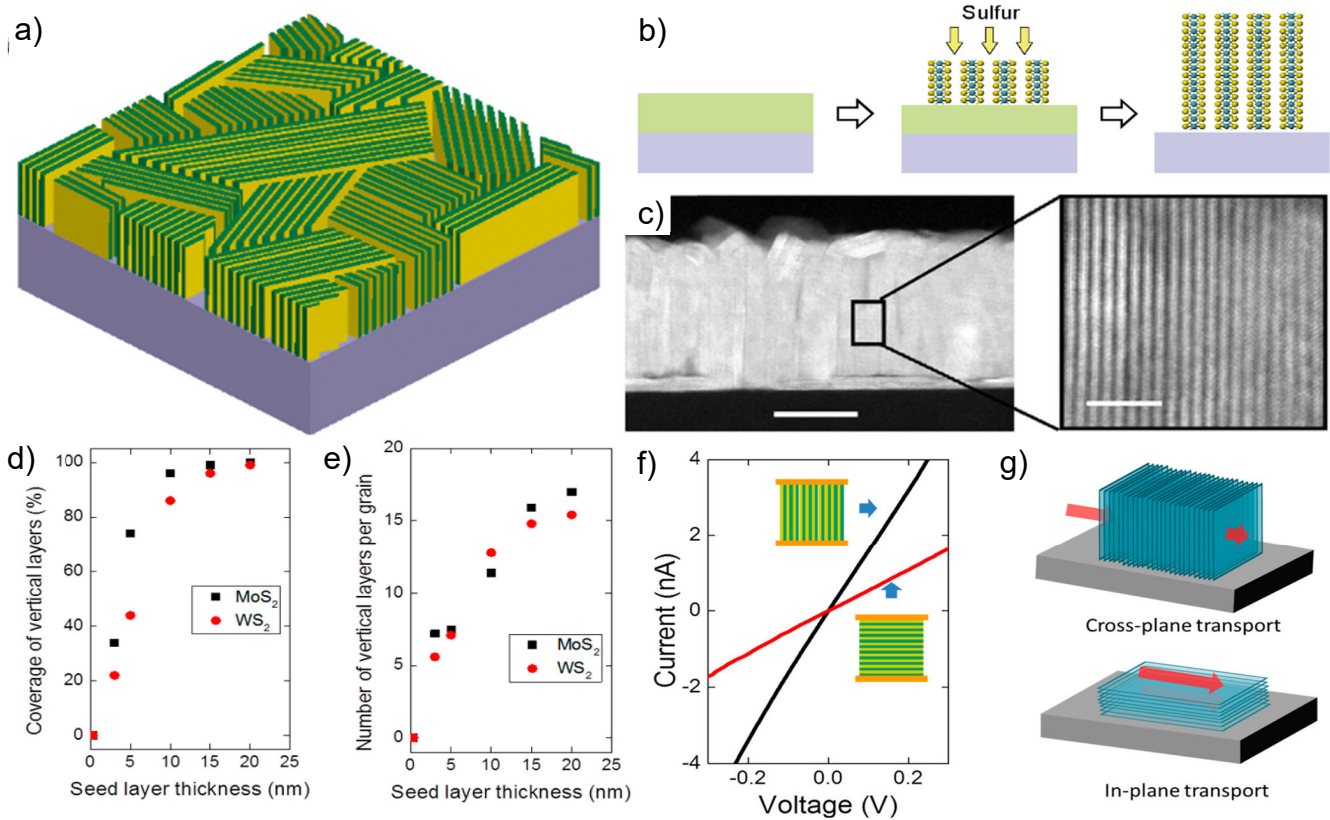

**Figure 6.** a) Schematic of a vertically aligned TMD film; b) Schematic of the growth process of a vertically aligned TMD film; c) Layer-resolved TEM image of vertically aligned TMD film; d) Plot of the measured percentage of vertically aligned grains and e) number of layers per vertical grain as a function of precursor thickness for 2SC $MoS_2$ and $WS_2$ films; f) Two-terminal I-V measurement of vertically aligned (black line) and horizontally aligned (red line) $WS_2$ films and g) schematic of the current pathway. a,b) Reproduced with permission.[147] Copyright 2013 American Chemical Society. c) Reproduced with permission.[61] Copyright 2023 Elsevier. d,e,f,g) Reproduced with permission.[145] Copyright 2014 American Chemical Society.

Growth of vertically aligned TMD films is not energetically favorable as the TMD edges have much larger surface energy than their basal planes.[148] Instead, the growth mode is expected to be kinetically driven. The reaction rate of formation of $MX_2$ is much larger than the crystallization rate. Meanwhile, the mobility of metal ions is low and the diffusion rate of chalcogen atoms through the van der Waals gaps of the growing film is high. As metal precursors undergo conversion, their lattices expand, providing stress in the lateral direction which can be relaxed by continued growth in the vertical direction.[61] The stress due to conversion as well as the initial roughness of the precursor film surface leads to the nucleation of both horizontal and vertical grains. As grains grow, high points and reactive edges on the film will collect chalcogens more readily. The resulting concentration gradient drives the diffusion of chalcogen atoms through the converting film. As the reaction rate is much faster than the TMD crystallization rate, the chalcogen atoms diffuse easily through the vdW gaps of the vertically aligned crystallites until they reach the metal precursor film where they rapidly react to form $MX_2$ and are slowly incorporated into the growing crystallites (Figure 6b).[31,147,149]

The density of vertically aligned layers in converted TMDs is highly sensitive to the thickness of the precursor layers.[145,147,150] Jung et al.[145] determined the ratio of vertical and horizontal grains through HR-TEM analysis and correlated the density of vertical grains with precursor film thickness for a fixed conversion process (See Figure 6d and 6e). At small precursor film thicknesses, they show that the fraction of vertically aligned grains grows linearly with greater precursor thickness. Using metallic Mo and W precursor films thicker than 10 nm, films become primarily aligned in the vertical orientation. The thickness dependent alignment of growing TMD films can be attributed to roughening of the film surface early in the deposition process which selectively favors vertically oriented grains with sufficient chalcogen pressure. Note, however, that this work probes precursor thicknesses down to 3 nm, as conversions of precursor films less than 1nm show 1-3 TMD layers which are highly aligned horizontally across a wide range of conversion conditions. Thus, at sufficiently low precursor film thicknesses, vertically aligned growth is not favorable.

Other sources of roughness can then be expected to induce more vertical alignment. Indeed, methods of reducing roughening such as use of single crystalline substrates,[97] increasing precursor smoothness,[126] increasing precursor purity,[126] prevention of metal agglomeration during conversion,[142] and passivation of the precursor surface by predeposition of a chalcogen layer[125] all have been suggested to reduce the tendency towards vertically aligned grains.

This has allowed the precursor thickness to act as a control knob for the orientation of converted TMDs. In contrast to the inert, smooth basal planes of 2D materials, the edges of these layered materials possess a high density of dangling bonds which are chemically active. These active sites can be used for chemical manipulation of material properties such as doping and functionalization.[151,152] Additionally, charge transfer at TMD edges is much faster than at the basal planes, making the active edges of these materials especially important for many catalytic reactions including hydrogen evolution reaction,[145,153–155] oxygen reduction reactions,[156] $CO_2$ reduction,[157] hydrodesulfurization,[153] etc. Further, the anisotropy of 2D materials extends to its optical and electronic properties. Figure 6f shows the measured current as a function of voltage through both vertically and horizontally aligned 2SC $WS_2$ films demonstrating that the in-plane electronic conductivity is much larger than the conductivity between layers. This dramatically changes the resistance between vertically and horizontally aligned TMD films and reduces the optical refractive index for light incident on vertically oriented films compared to horizontal ones.[145] Using precursor thickness to control orientation of converted TMDs, Jung et al.[145] pattern their precursor films with stripes of different thicknesses. Upon conversion, these produce a grating with alternating regions of horizontally aligned regions and vertically aligned regions which are confined to the areas of thin and thick precursor, respectively.

With a fixed precursor condition, one can also tune the reaction kinetics primarily through controlling the chalcogen partial pressure and the conversion temperature. Maintaining a high chalcogen partial pressure during conversion has been shown to promote vertical alignment using elemental powders (by controlling the crucible temperature).[158,159] High chalcogen content results in larger concentration gradients which promotes chalcogen diffusion. Thus, using a high chalcogen content enables rapid conversion of films with uniform, vertically aligned TMD layers.[147] Although vertically oriented 2SC TMD films have been fabricated using both elemental powders and hydride gasses, the differences between the effects they have on orientation is not clear and requires further experimentation.

Compared to the chalcogen content, the effect of temperature on film orientation is more subtle as increasing temperature increases both the chalcogen diffusion rate as well as the TMD crystallization rate. It has been shown that for fixed chalcogen content, sufficiently large temperatures promotes horizontal grain growth; indicating that the crystallization rate dominates growth kinetics over chalcogen diffusion.[145]

## 6. Heterostructures, Superlattices, and Encapsulation

One advantageous property of 2D materials is the ease and versatility of vertically stacking layers to form artificially engineered bulk materials.[160–162] Without out-of-plane dangling bonds, 2D materials can be combined to create 3D structures without inducing strong carrier scattering and bypass epitaxial lattice-matching requirements typical for synthetic heterostructures. This leaves the electronic properties of constituent layers largely distinct with interlayer interactions acting as a perturbation. Combining of disparate materials into one heterostructure enables unique control of properties, including tunable bandgap values,[16] enhanced light-matter coupling,[163] improved electronic performance, and enhanced catalytic activity.[164] As such, heterostructures provide exciting opportunities to probe the fundamental electronic interactions at the interfaces of 2D/2D and 2D/3D systems (and emergent phenomena)[165–173] giving control to tune material properties and design atomically thin device structures which take advantage of the properties of individual layers.[160,162,174,175]

To date, the primary and most widely used method of producing heterostructures is by mechanically or wet-transfer stacking of 2D material flakes or thin films. Through either liquid or dry transfer processes,[176–179] 2D flakes or even large area films can be assembled into arbitrary-thicknesses with precise control of twist angle, stacking configuration, and architectures.[180] For example, through careful design of the repeating unit cell in a vdW superlattice, large area TMDs/h-BN superlattices can enable >90% narrow band absorption in less than 4 nm of active layer excitonic absorber.[163] However, this process is time consuming and is often plagued by contamination at the interface between materials. Thus, limited heterostructure quality, repeatability, and scalability have impeded the development of heterostructure-based devices.

Direct growth of 2D heterostructures and superlattices with a high degree of control and process scalability remains an outstanding challenge for the field. Recent advancements in MOCVD growth have shown to directly grow multiple layer alternating monolayer films of TMDs to form complex 3D nanomaterials, but the low nucleation density results in synthesis processes extended over many days.[181] The 2SC process can provide an alternative, rapid, and scalable method with repeatable interface cleanliness to heterostructure synthesis and their vertical superlattices (repeating units of a heterostructure) by alternating precursor deposition and conversion.[136,182,183] Here, the precursor deposition technique should be carefully chosen and deposition energies tuned to reduce damage caused to the underlying TMD layers. Motala et al.[61] recently reported a two-step method of producing heterostructures and superlattices where an alternating stack of two precursor metals (eg. Mo/W/Mo/W) is first deposited by magnetron sputtering followed by conversion to selenides using $H_2Se$ (Figure 7a and 7b). This results in a TMD heterostructure in controllable formation of either alternating layered units ($MoSe_2/WSe_2/MoSe_2/WSe_2$) or an alloyed structure, depending upon the annealing temperature (Figure 7c). These superlattice structures are shown to enhance the optical performance over individual TMD layers due to the modified band structure and enhanced light trapping in the repeating structure (Figure 7d and 7e). However, one important consideration using this heterostructure fabrication process is the potential intermixing of metal ions into adjacent layers (shown in Figure 7c) as the authors show that at elevated temperatures, the metal layers intermix to form a uniform alloy rather than distinct layers. To combat the challenge of intermixing of layers, Zhou et al.[62] develop a sequential deposition scheme where subsequent layers are converted at lower temperatures (Figure 7f) in order to avoid decomposition of the underlying layer. This method succeeds in producing distinct layers down to the monolayer (shown in Figure 7g) of materials down to the monolayer limit. High quality few-layer $MoS_2/WS_2$ heterostructures can also be synthesized via 2SC which show promise for various optoelectronic applications.[136]

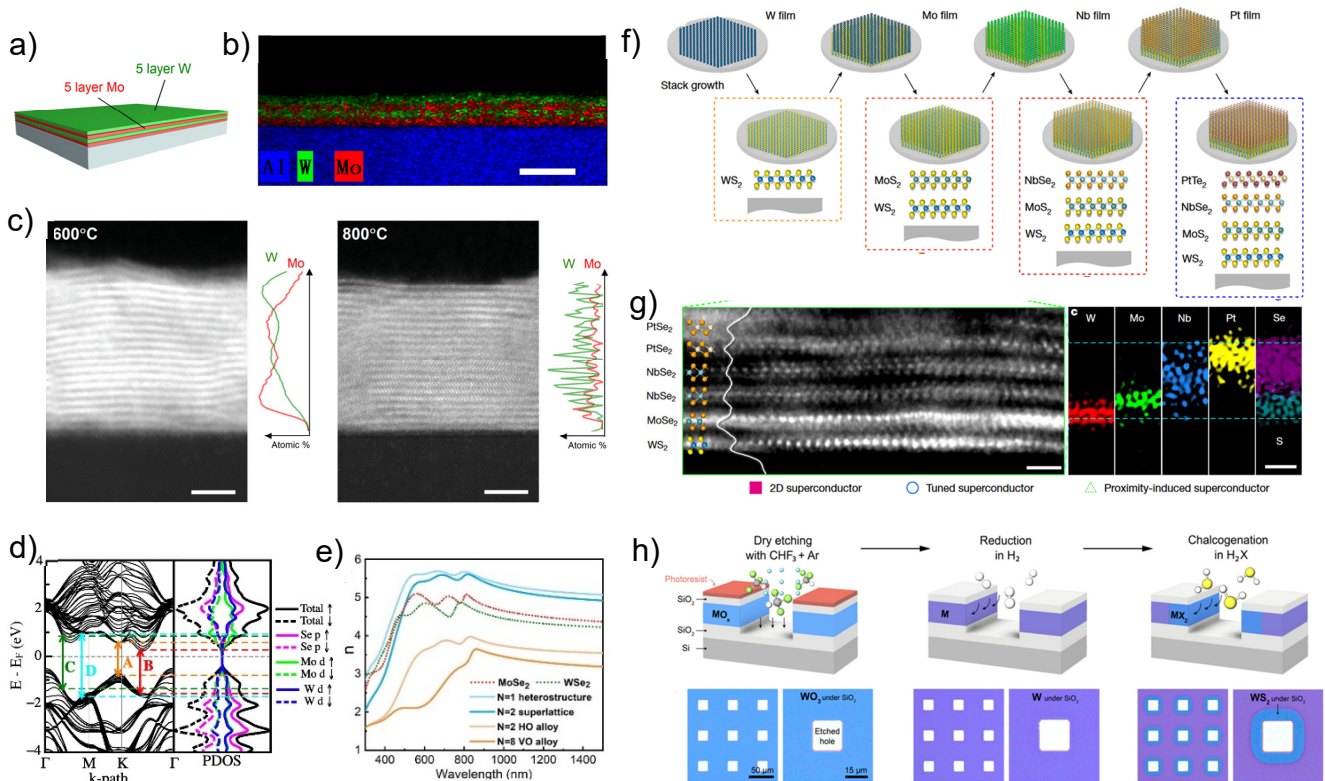

**Figure 7.** a) Schematic and b) cross-sectional TEM image of an alternating stack of ultrathin Mo and W precursor layers; c) TEM image and EDS signal measured from a superlattice of MoSe$_2$ and WSe$_2$ converted at 600 °C (left) and a fully alloyed film when converted at 800 °C (right); d) DFT calculated band structure of a MoSe$_2$/WSe$_2$ superlattice with low energy optical transitions labelled; e) Measured index of refraction of a 2SC MoSe$_2$/WSe$_2$ heterostructure, superlattice (N=2), and horizontal and vertically aligned alloyed films; f) Schematic depiction of the sequential deposition and conversion method of producing 2SC heterostructures; g) Cross-sectional TEM image and EDS map of a sequentially deposited heterostructure; h) Schematic (top row) and optical images (bottom row) of method of producing encapsulated 2SC films by conversion of a buried precursor film. a,b,c,d,e) Reproduced with permission.[61] Copyright 2023 Elsevier. f,g) Reproduced with permission.[62] Copyright 2023 Springer Nature. h) Reproduced under terms of the CC-BY license.[75] Copyright 2019 IOP Publishing Ltd.

Beyond 2D/2D heterostructures, 2D/3D mixed dimensional heterostructures can leverage the benefits of the 2SC method. For instance, mechanical stacking of 3D materials onto 2D layers can protect TMD films from degradation through interactions with ambient air and additionally can be used as a metallic contact or dielectric barrier for device fabrication. The 2SC technique awards a simpler route to encapsulation as chalcogen atoms can readily diffuse into the metal precursor film through the exposed edges of the metal/encapsulant stack. This effect was demonstrated by Kemelbay et al.[75] by encapsulating a metal precursor film in amorphous $SiO_2$ prior to the conversion stage. They showed that the chalcogen atoms diffuse into the buried precursor film, converting it into an encapsulated TMD. For example, Pace et al.[60] transfer a large area layer of hBN onto an air-sensitive 2SC grown $MoTe_2$ film, preserving the properties of the film from approximately 3 hours to greater than one month by protection from oxidation.

Although the development of 2SC heterostructures is growing as quickly as the 2SC technique itself, accurate, rapid, and non-destructive characterization of heterostructure interfaces across large-area films is lacking. As the quality of the van der Waals interface between materials has a large impact on heterostructure performance, characterization of these interfaces requires elemental differentiation at angstrom-scales. Toward this goal, TEM coupled with EDS has shown to be the only suitable characterization technique to date for truly understanding atomic layer chemistry. Indeed, TEM and EDS have been used to demonstrate both elemental separation and intermixing at heterostructure interfaces under different process conditions.[61,62] Some caveats should be noted when using these techniques for heterostructure characterization. Firstly, cross-sectional TEM requires ion-beam milling and conditioning of a thin section of a heterostructure film. The ion-beam is known to impart significant energy to the lattice near the milled surface which can introduce large-scale distortion of the layers as well as atomic-scale disorder which may introduce atomic mixing. Second, EDS measures the elemental distribution throughout the volume of the measured thin section of a sample, therefore, distortion of individual layers, section thickness, and alignment of the measuring electron beam can introduce significant uncertainty in the measured profile. Finally, TEM and EDS are limited to an extremely small sampling volume compared to the wafer-scale heterostructure film. Combined with the difficulty of sample preparation, TEM+EDS is typically performed only at a few locations on the sample and images are often chosen to represent the highest quality regions of a film. Thus, this technique is often not a good representation of the quality of a heterostructure film across wafer scales. More sampling by TEM and EDS as well as other characterization techniques, such as scanning probe based methods,[184,185] is needed to provide a more accurate description of wafer-scale heterostructure uniformity.

## 7. Application opportunities

The advancement in large scale, reliable synthesis techniques for 2D materials have arguably been the most compelling development towards enabling real-world applications. 2SC, MOCVD, CVD, PVD, liquid and mechanical exfoliation techniques are now transitioning, or in the process of transitioning, into major industries and material suppliers. The 2SC technique described herein, has demonstrated several key advantages for the growth of high quality 2D materials in electronics, sensing, energy, optics and optoelectronic applications (Figure 8). Large scale synthesis of bilayer $MoS_2$ transistors synthesized through conversion of ALD-synthesized $MoO_3$ have shown to produce field effect transistors with exceptional performance even on flexible substrates. The transistors arrays demonstrate field effect mobilities up to 55 $cm^2/Vs$, subthreshold slope down to 80 $mVdec^{-1}$ and on/off ratios of $10^7$ in a process amenable to 6-inch wafer processing.[80] Similarly, highly uniform monolayer $MoS_2$ FETs have been produced on 2" sapphire through sulfurization of ALD $MoO_3$ with mobilities as high as 30 $cm^2V^{-1}s^{-1}$, ON/OFF current ratio of about $10^7$, and hysteresis as low as 0.4 V.[86] Vertical memristor devices utilizing $MoS_2$ films from sulfurized $MoO_3$ show nonvolatile resistive switching with $I_{on}/I_{off}$ ratios of $10^4$ at low bias voltages of $\pm 1$ V.[36] The resultant TMD thickness using this technique is very uniform, crucial for improving gate control and reducing short-channel effects. Modulation of the TMD thickness can be finely controlled through the precursor metal or metal oxide layers. Moreover, several examples of the 2SC process demonstrate conversion temperatures below 400 ºC, a crucial benchmark to realizing back-end-of-line (BEOL) silicon processing temperatures, while still realizing electronic and optical properties comparable to materials grown by methods requiring significantly higher temperatures.

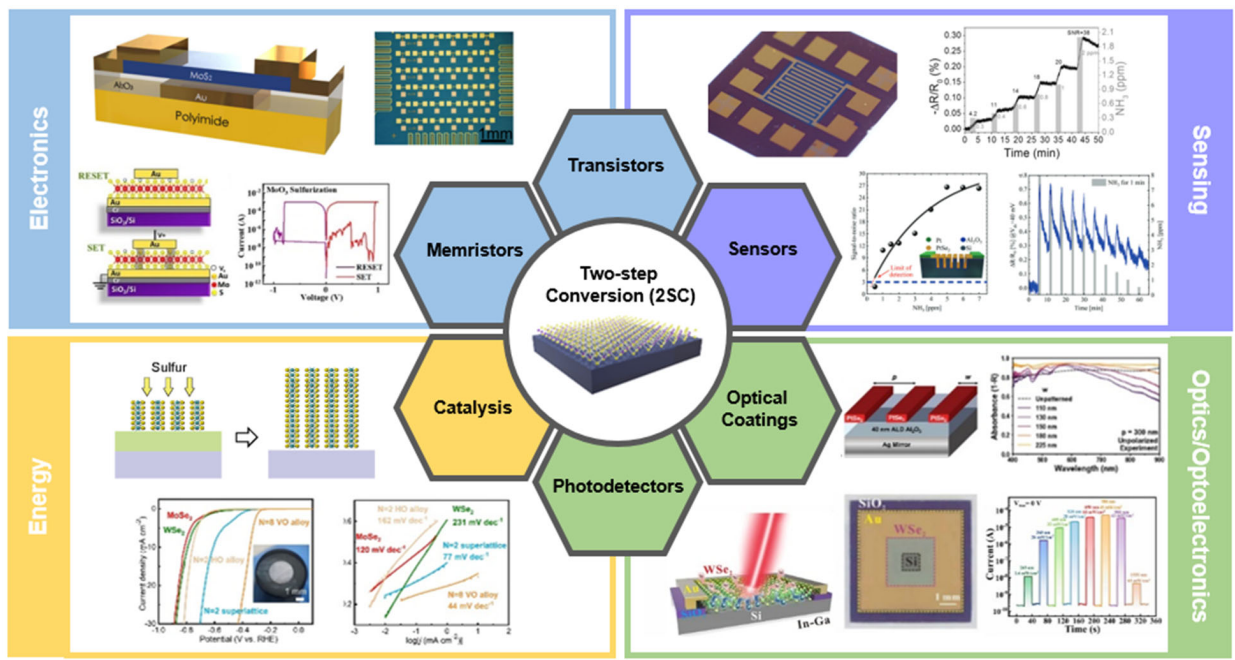

**Figure 8:** Diverse application opportunities of TMDs synthesized by 2SC. Reproduced under terms of the CC-BY license.[80] Copyright 2023 Springer Nature. Reproduced with permission.[36] Copyright 2023 American Chemical Society. Reprinted with permission.[186] Copyright 2013 Wiley-VCH. Reproduced with permission.[83] Copyright 2021 Wiley-VCH. Reproduced with permission.[187] Copyright 2022 Wiley-VCH. Reprinted with permission.[140] Copyright 2023 American Chemical Society. Reprinted with permission.[147] Copyright 2013 American Chemical Society. Reproduced with permission.[61] Copyright 2023 Elsevier.

The combination of excellent semiconducting properties, low surface-to-volume ratio, and controllable active sites have propelled 2D materials to be a prime candidate for future low cost, integrated, highly sensitive vapor or liquid sensors for environmental and health monitoring. The advantages of electronic applications using 2D materials synthesized by the 2SC approach can be extended to sensor applications, where BEOL integration and deposition on a variety of substrates will be vitally important. Also critical in this case is the ease and speed of manufacturing. For instance, as a response to the SARS-CoV-2 pandemic, billions of at-home diagnostic tests had to be manufactured in an alarmingly fast pace.[188] In the case of the 2SC synthesis approach, it is feasible to envision rapid wafer production, where a combination of scalable metal/metal oxide deposition techniques combined with low temperature conversion could produce large wafers on the scale of minutes. As 2D materials have shown some of the most impressive sensing performance in response to the SARS-COV-2 outbreak,[189,190] rapid production of low cost, scalably produced 2D wafers will be important in the future. This is also true for vapor sensing, where the detection of harmful chemicals using 2D semiconductors can push beyond current state-of-the-art sensors. For instance, a simply fabricated $MoS_2$ interdigitated sensor converted from a sputtered Mo film showed a limit of detection approaching 50 ppb for $NH_3$, orders of magnitude lower than the current safety threshold.[186] Highly conformal $PtSe_2$ sensor devices, fabricated from the conversion of ALD-deposited Pt, were also demonstrated with a theoretical limit of detection of 370 ppb in a highly scalable low temperature (450 °C) process.[83]

Similar to electronic and sensing devices, optical coatings and photodetectors can also benefit substantially from the low temperature, versatile substrate accommodation, and high-quality films realized by the 2SC approach to TMD synthesis. Synthesis of $PtSe_2$ coatings through conversion of ultrathin Pt films have demonstrated extraordinarily large refractive index and extinction, which are governed by interband optical transitions. This large, broadband refractive index can be exploited to create superabsorbers with strong absorption (>87%) in the 400-900 nm range for films of only 18 nm.[187] $PtSe_2$-based Photodetectors can easy be fabricated on conformal surfaces and demonstrate high responsivity (241 $mVW^{-1}$) with BEOL-compatible processing.[83] Other wafer-scale, integrated photodetectors devices using $PtTe_2$[132] and $WSe_2$ have shown promising results comparable to TMDs growth by other techniques. The ability to create heterostructures using the 2SC approach can be a viable route towards tunable and diverse optical. For example, creation of complex heterostructures and alloys through the 2SC method have been shown to directly influence band structure, bandgap and absorption for alternating few layer $WSe_2$/$MoSe_2$ sequential heterostructures.[61]

One extremely unique feature of the 2SC approach, highlighted in prior sections, is the ability to control layer orientation by modulating precursor thickness and conversion conditions. With vertically-oriented films featuring TMD layers growing perpendicular to the growth substrate, edge-terminated films can provide maximal sites for HER and catalytic activity.[155,191] Moreover, the vertical structure ensures efficient charge flow along the TMD layer from the conductive support to the active surface sites, reducing ohmic losses and ensuring better performance.[147] A vertically oriented $Mo_{1-x}W_xSe_2$ alloyed structure synthesized via the conversion of Mo/W sequential layers demonstrated an extremely low Tafel slope of 44 mV/dec, the lowest among Tafel slopes reported for TMD films.[61] This orientation control can not only control active sites for catalyisis, but lead to vastly different electronic, optical, and thermal properties for future diverse applications.

## 8.    Outlook

It is clear that for applications demanding direct growth of a variety of TMD materials on diverse substrates, at low temperatures, and with exceptional quality, the 2SC synthesis approach for monolithic 2D materials and their heterostructures can offer significant advantages over other established methods even beyond those outlined herein. Utilization of highly conformal precursor deposition techniques such as highly-ionized sputtering and ALD enable 2D materials synthesis with uniform thickness and properties on high aspect ratio trenches or other features. Growth rates for precursors are generally much higher than direct application of TMD films via CVD, MBE, ALD, and MOCVD, and precursor-to-TMD conversion temperatures lower, yielding materials and with much higher structural quality than directly sputtered TMD films. Pre-patterning of desired phases and ease of application for low melting point and noble metal TMDs are also facilitated by embodiments of the 2SC process. Even in comparison to other synthesis techniques, properties including electron mobility, absorption, catalysis, and sensing are on par with more conventional strategies.

While the opportunities for 2SC growth of TMDs are vast, there remains more work to be done. An improved understanding of precursor conditions (including grain size, oxidation state, and doping) and the effect upon properties of resultant TMDs is a necessary step to understanding critical mechanisms for the conversion process. For example, increasing TMD grain sizes is expected through careful control of precursor crystal structure. Modulating the kinetics of the conversion process can realize far-from-equilibrium materials which may unlock new crystalline phases and chemistries yet to be explored, especially for heterostructures.[192] More efforts in developing reliable synthesis of monolayer materials and studying the interface between the semiconductor and substrate will be important for future applications. Finally, efforts to further reduce conversion temperatures, especially for BEOL processing below 400 °C, by generating alloys, using plasma enhancement processes, incorporating catalysts, and by dynamic vacuum annealing will be vital for easy incorporation with relevant technologies desperate to leverage the exciting 2SC synthesis method.[77,192,193]

## Acknowledgements

The authors gratefully acknowledge support from the Air Force Office of Scientific Research Grant #FA9550-24RYCOR011.